\documentclass[epj]{svjour}
\usepackage{amsmath}
\usepackage{amsmath}

\usepackage{latexsym}
\usepackage{graphics}
\begin{document}
\title{Lorentz violating inflation and the Swampland}
\author{Oem Trivedi\inst{1}
\thanks{oem.t@ahduni.edu.in}%
}                     
%
%
\institute{School of Arts and Sciences, Ahmedabad University,Ahmedabad 380009,India}
\date{Received: date / Revised version: date}
%
\abstract{The swampland conjectures from String theory have had very interesting implications for cosmology and particularly for Inflation. It has been shown that the single field inflationary models in a GR based cosmology are in unavoidable tensions with these conjectures, while these single field models can still be consistent in certain non-trivial inflationary regimes. So it becomes interesting to see whether there is a way to overcome the issues of the swampland and single field inflation in an essentially GR based cosmology. We show that this can indeed be the case and for this, we consider a certain type of Lorentz violating inflationary scenario. We work out the swampland bounds for Lorentz violating inflationary models after which show that inflationary models which would have had otherwise serious tensions with these conjectures in a usual GR based scenario, can be very tranquil with the criterion in this regime. For this we take examples of Higgs inflation, radion gauge inflation and Spontaneous symmetry breaking inflation and show that the bounds imposed by the swampland on the Lorentz violating parameter can be easily satisfied in these models.}

\maketitle

\section{Introduction}

   The concept of cosmic inflation has been extremely effective in explaining different properties of the early universe \cite{starobinskii1979spectrum,sato1981first,guth1981inflationary,linde1983chaotic,linde1995quantum} . Various satellite observations have consistently confirmed inflationary estimates for the early universe, and the most recent evidence from the Planck experiment continues this trend \cite{aghanim2020planck,akrami2020planck1,aghanim2018planck,akrami2018planck} . Furthermore observational evidence supports a wide range of inflationary models, a lot of which are inspired by radically different contexts ranging from modified gravity theories to quantum gravitational  \cite{martin2014encyclopaedia,martin2014best,wands2008multiple,berera1995warm,dvali1999brane,alexander2013chern,kanti2015gauss,oikonomou2021generalizing,Odintsov:2019clh} . However, traditional single field models (which some refer to as "supercooled inflationary models \cite{berera1995warm}") have a lot of empirical support and are still widely used in theoretical studies.
	
   In recent years, a lot of effort has gone into developing a “Theory of Everything,” and String Theory is probably the most well-known candidate for such a paradigm. Since string theory is presented in such a vivid way, it is reasonable to expect it to have far-reaching consequences for cosmology.
	As a result, there is a large and diverse body of literature that has looked into the cosmological consequences of string theory. The extremely high number of possible vacua states that string theory makes, which can go as high as $\mathcal{O}(10^{500})$ , is one of the many cosmologically interesting aspects of the theory and all these possible vacua constitute the "Landscape" of string theory. The question of which class of low energy EFT's are actually compatible with string theory then becomes a very interesting one. In order to address this issue, Vafa coined the word "Swampland," which refers to a class of low-energy effective field theories that could not be have a consistent UV completion with regards to the string theoretic paradigm. Furthermore, a number of field theoretic UV completion criterion from string theory known as the "swampland conjectures" have been proposed in recent years to classify whether a given regime is in the swampland or not. Many people consider string theory to be a feasible quantum gravity framework, so if a low-energy EFT meets these criteria, it might also get along with a self-consistent quantum gravity theory. Although quite a number of swampland criterion have been proposed in recent times, like the completeness \cite{polchinski2004monopoles} and cobordism \cite{mcnamara2019cobordism} conjectures,the distance and the dS conjectures have been shown to quite clearly have the most striking implications for cosmology of all these conjectures. These two conjectures can be described as follows:  
	\\
	\\
	$ 1 $ : Swampland Distance Conjecture : This conjecture limits the field space of validity of any effective field theory \cite{ooguri2016non} . This sets a maximum range traversable by the scalar fields in an EFT as \begin{equation}
	\frac{\Delta \phi}{m_{p}} \leq d \sim \mathcal{O} (1)
	\end{equation} 
	where $ m_{p} $ is the reduced Planck's constant, d is some constant of $ \mathcal{O} (1) $ , and $\phi$ is the Scalar Field of the EFT.
	\\
	\\
	$ 2 $ Swampland De Sitter Conjecture : This Conjecture states that it is not possible to create dS Vacua in String Theory \cite{obied2018sitter}. The conjecture is a result of the observation that it has been very hard to generate dS Vacua in String Theory \cite{dasgupta2019sitter,danielsson2018if}( While it has been shown that creating dS Vacua in String Theory is possible in some schemes ,like the KKLT Construction \cite{kachru2003sitter}). The Conjecture sets a lower bound on the gradient of Scalar Potentials in an EFT , \begin{equation}
	m_{p}  \frac{| V^{\prime} |}{V} \geq c \sim \mathcal{O} (1)
	\end{equation} 
	where c is some constant of $ \mathcal{O} (1) $ , and V is the scalar Field Potential. A " refined " form of the Swampland De Sitter Conjecture places constraints on the hessian of the scalar potential (a finding which first appeared in \cite{garg2019bounds} and later in \cite{ooguri2019distance} ) and is given by \begin{equation}
	m_{p}^{2}   \frac{ V^{\prime \prime}}{V} \leq - c^{\prime} \sim \mathcal{O} (1)
	\end{equation}
	These criterion have intriguing implications for cosmology, especially single-field inflation. Considering the data on inflation, it was shown that single field inflation in a GR based cosmology is incompatible with these conjectures \cite{kinney2019zoo}. A lot of effort has gone into resolving this dispute \cite{geng2020potential,scalisi2019swampland,ashoorioon2019rescuing}. However, it has been shown that if the background cosmology for single field inflation is not GR based, then this inflationary regime can still satisfy the swampland criterion\cite{lin2019chaotic,odintsov2020swampland,yi2019gauss,trivedi2020swampland,oikonomou2021rescaled}. Multi field inflationary models have also been shown to be quite consistent with the conjectures even in GR based paradigms \cite{bravo2020tip}. It's also worth noting that in both GR and non-GR based cosmologies, the warm inflation paradigm has been shown to be very compatible with the swampland criterion even for single field models \cite{motaharfar2016warm,motaharfar2019warm,das2019warm,das2020runaway} The recently proposed “Trans Planckian Censorship Conjecture” (TCC)\cite{bedroya2019trans} is another swampland conjecture that has sparked a lot of interest in inflationary cosmology . Models with tachyonic scalar fields as the inflaton  can also be consistent with the conjectures \cite{mohammadi2020warm,trivedi2021rejuvenating}. Single field GR based inflationary models can only be compatible with the TCC if they are extremely fine tuned \cite{bedroya2020trans}, which is ironic given that inflation was invented to solve the fine tuning problem of traditional big bang cosmology. Much further work has been done to clarify the TCC's problems with single field inflation, with the overall picture appearing to be that the TCC is more comfortable with non-standard inflationary regimes than usual single field models \cite{bernardo2020trans,mizuno2020universal,brahma2020trans,dhuria2019trans,brandenberger2020strengthening,schmitz2020trans,guleryuz2021trans}. What one can hence understand from the current literature on the swampland and inflation is that one is not well placed to expect conventional(cold) single field models to be on amicable terms with these conjectures. Hence, it would optimistic for one to think that non-trivial modifications to cold single field models based in a GR cosmology could lead to these paradigms being consistent with the conjectures. \\ \\
	One such non trivial modification to single field inflationary models can be found by investigating such regimes in a Lorentz violating background. Several inflationary regimes in Lorentz-violating cosmological scenarios have been studied in the recent times \cite{gasperini1985inflation,lim2005can,zuntz2008constraining,armendariz2010primordial,kanno2006Lorentz,avelino2009impact,donnelly2010coupling,almeida2017cosmology}. Gasperini \cite{gasperini1985inflation} proposed that the primordial period of rapid expansion of the Universe may be achieved if the gravitational interactions were characterised by a non locally Lorentz invariant theory at some extremely early epoch.The authors in \cite{donnelly2010coupling} examined a Lorentz-violating inflationary theory based on an Einstein-aether theory and a scalar field Lagrangian. Such scenarios have also been used to understand dark energy, where it has been demonstrated that violating the Lorentz invariance creates Lagrangians capable of driving the universe's current acceleration \cite{blas2011technically,audren2013cosmological}. The former and latter theories, as a result, deal with Lorentz symmetry violation at small and large distances, respectively \cite{jacobson2008einstein}. In \cite{almeida2017cosmology}, it was shown that a time-like Lorentz violating background can produce sufficient inflation whilst also providing an explanation for the current dark energy epoch. Another interesting point with Lorentz violations is that Lorentz violations can be found in certain quantum gravity constructions \cite{Jacobson:2004qt,collins2006lorentz,Mavromatos:2007xe,li2011background}. Lorentz violation during inflation generates some non standard changes in the usual inflationary dynamics and could hence be an interesting regime with regards to the swampland.We also want to emphasize an important point, which is that the swampland conjectures have been aimed to be criterion for the underlying quantum gravity as a whole, even though it is based in string theory. Hence we are not under any restriction to only consider some particular form of Lorentz violating cosmology which is suited or based in some specific quantum gravity approach. The goal of this paper here is to understand whether Lorentz violations during inflation (and more generally in the early universe ) could possibly be appreciated by the underlying theory of quantum gravity and for achieving that, we use the swampland criterion.
	\\
	\\
	In the next section,we briefly describe about a particular kind of Lorentz violating cosmology, while in section III we discuss the swampland status quo of (cold) single field inflation in that regime. In section IV, we will consider 3 different inflationary models and discuss which of them could be consistent with the swampland on the basis of the groundwork made in Section III. We summarize our work in section V with some concluding remarks about the overall picture of the early universe painted by the swampland so far.
	\section{Basics of the Lorentz violating cosmology}
	
	We will be working in a time-like Lorentz violating cosmological background here.A purely time-like background can be defined by a tiny subset of Lorentz invariant violating operators that preserve rotational invariance, which is why one would use a time-like Lorentz operator here \cite{passos2017Lorentz} . Furthermore as Kostelecky and Mewes describe in \cite{kostelecky2009electrodynamics} , the Cosmic Microwave Background becomes a natural choice of a preferred frame in this scenario.	We start off with a Lagrangian in a 3+1 dimensional background describing a scalar field ( which will eventually be the inflaton) coupled to a Lorentz violating background (similar to the one here \cite{almeida2017cosmology}) \begin{equation}
	\mathcal{L} = \Bigg( \frac{R}{2 \zeta^{2}} - \frac{1}{2} \left( g{{_{\mu}}{_{\nu}}} + \xi_{1} k^{1}{{_{\mu}}{_{\nu}}}  \right) \partial^{\mu} \phi \partial^{\nu} \phi - V(\phi) \Bigg) \sqrt{-g}
	\end{equation}
	where $ \zeta^{2} = 8 \pi G $ and $V(\phi) $ is the inflaton potential while $k^{i}{{_{\mu}}{_{\nu}}}$ are time-like tensors which can even couple to other fields in a more general Lagrangian. It does so in a way such that the only non-zero component is $ k^{i}{{_{0}}{_{0}}} $ , which means \begin{equation}
	k^{i}{{_{\mu}}{_{\nu}}} = \begin{pmatrix}
	-\beta_{i} & 0 \\ 
	0 & 0 
	\end{pmatrix}
	\end{equation}
	where the tensor couples to the scalar field via the coupling $ \xi_{1} >0 $. This kind of a theory has been touted to be able to explain both the current expansion of the universe as well the inflationary phase \cite{almeida2017cosmology}. It can do so if  $ \beta_{1}\to 0 $ for short distances, which allows for all the inflationary dynamics to be controlled only by the inflaton field while if $ \beta_{1} \to -1 $ for large distances, the Lorentz violating regime can also explain dark energy. We will, however, just be discussing about the inflationary phase in our analysis in the next section. 
	For studying the cosmological properties of the Lagrangian described above, we can take the metric to be of the FLRW form \begin{equation}
	ds^{2} = -dt^2 + a(t)^{2} d\boldsymbol{x}^{2} = g{{_{\mu}}{_{\nu}}} dx^{\mu} dx^{\nu} 
	\end{equation}
	The background fields responsible for the Lorentz violating factor would change the metric as \begin{multline}
	d\overline{s}^{2} = \overline{g}{{_{\mu}}{_{\nu}}} dx^{\mu} dx^{\nu} =  \left( g{{_{\mu}}{_{\nu}}} + \xi_{1} k^{1}{{_{\mu}}{_{\nu}}}  \right) dx^{\mu} dx^{\nu} = \\ -\left(1 + \xi_{1} \beta_{1} \right) dt^{2} + a(t)^{2} d\boldsymbol{x}^{2}
	\end{multline}
	here one can then go on to write an effective velocity for the scalar field as \begin{equation}
	v = \sqrt{1 + \xi_{1} \beta_{1} } c
	\end{equation}
	where c is the usual speed of light. One can then also define a relationship between the Lorentz violating parameter and the redshift z as \cite{almeida2017cosmology} \begin{equation}
	\xi_{1} \beta_{1} = - \frac{C_{1}}{z(z+2) + C_{1}}
	\end{equation}
	where $ C_{1} $ is a constant. This relation becomes very important when one wants to realize both the inflationary phase and the dark energy phase with the same inflation field but as we are currently interested only in the former phase we will not explore the redshift relation further. 
	If one considers a General Relativity based cosmological scenario, then the Friedmann equation will be remain the same as they are in the usual GR sense. Hence, the Friedmann equation remains as \begin{equation}
	H^{2} = \frac{\rho}{3 m_{p}^{2}}
	\end{equation} 
	where we are working in $ m_{p} = \sqrt{ \frac{1}{8 \pi G} } $ units. However, the energy and pressure densities will not remain as they usually are in GR. The energy-momentum tensor can be written here as \begin{equation}
	T{{_{\mu}}{_{\nu}}} = \partial_{\mu} \phi \partial_{\nu} \phi + g{{_{\mu}}{_{\nu}}} \mathcal{L}_{\phi} 
	\end{equation}
	where $ \mathcal{L}_{\phi} = - \frac{1}{2} \left( g{{_{\mu}}{_{\nu}}} + \xi_{1} k^{1}{{_{\mu}}{_{\nu}}}  \right) \partial^{\mu} \phi \partial^{\nu} \phi - V(\phi) $. 
	Now we can write down the pressure and energy densities, given that our field configurations are homogeneous and so $ \phi = \phi (t) $, we have the energy and pressure densities as \begin{equation}
	\rho_{\phi } = \frac{1}{2} (1 + \xi_{1} \beta_{1} ) \dot{\phi}^{2} + V(\phi)
	\end{equation} 
	\begin{equation}
	p_{\phi } = \frac{1}{2} (1 - \xi_{1} \beta_{1} ) \dot{\phi}^{2} - V(\phi)
	\end{equation}
	Finally, the equation of motion for the scalar field is found out to be \begin{equation}
	\ddot{\phi} + 3 H \dot{\phi} + \frac{V^{\prime}(\phi)}{1 + \xi_{1} \beta_{1}} = 0 
	\end{equation}
	This completes our groundwork for the formulation of inflation in this scenario. In the next section we will be discussing about the different inflationary parameters in this scenario and consider the possibility of this regime being swampland consistent.
	\section{Swampland implications for Lorentz violating inflation }
	
	In the inflationary phase, the energy density $ \rho $ is dominated by the inflaton energy density $ \rho_{\phi } $ and so one can take $ \rho \sim \rho_{\phi } $ during inflation. Hence, the Friedmann equation (10) during inflation takes the form \begin{equation}
	H^{2} = \frac{1}{3 m_{p}^{2}} \left( \frac{1}{2} (1 + \xi_{1} \beta_{1} ) \dot{\phi}^{2} + V(\phi)  \right)
	\end{equation}
	From now on wards in the paper, we define the Lorentz violating parameter to be $ \kappa = \xi_{1} \beta_{1} $. Now during inflation, $ \dot{\phi}^{2} << V(\phi) $ which corresponds to the slow roll condition. Applying this criterion here, we find the Friedmann equation to be \begin{equation}
	H^{2} = \frac{V(\phi)}{3 m_{p}^{2}} 
	\end{equation}
	One can note that this is the same form that the Friedmann equation has for all sorts of inflationary models in a GR based cosmology, which are Lorentz-abiding regimes. Hence, the Friedmann equation remains unchanged even after factoring in Lorentz violations in this case. Now, we cast our attention to the field equation for the inflaton (14). Another slow roll criterion during inflation is that $ \ddot{\phi} << 3H \dot{\phi} $. Considering this, we can write the field equation to be \begin{equation}
	3H \dot{\phi} + \frac{V^{\prime}(\phi)}{1 + \kappa} \approx 0 
	\end{equation} 
	Now we begin to see effects of Lorentz violation in the inflationary dynamics, as the Lorentz violating parameter plays an important role in the field equation during inflation.
	Now lets find expressions for an important part of the inflationary setup, the slow roll parameters. The $ \epsilon$ slow roll parameter is given by \begin{equation}
	\epsilon = -\frac{\dot{H}}{H^{2}}
	\end{equation} 
	Using the Friedmann equation (16) and the Field equation (17), we can write the $ \epsilon $ potential slow roll parameter from its usual definition as \begin{equation}
	\epsilon = \frac{m_{p}^{2}}{2 (1 + \kappa)} \left( \frac{V^{\prime}}{V}  \right)^{2}
	\end{equation}
	And further, we can find the $\eta$ potential slow roll parameter similarly as \begin{equation}
	\eta = \frac{m_{p}^{2}}{1 + \kappa} \frac{V^{\prime \prime}(\phi)}{V(\phi)}
	\end{equation}
	After this, one can then find out the e-fold number during Inflation as \begin{equation}
	N = \int_{t_{i}}^{t_{f}} H dt 
	\end{equation}
	This can then be written in terms of $\phi$ as \begin{equation}
	N = \int_{\phi(t_{i})}^{\phi(t_{f})} \frac{(1 + \kappa)}{m_{p}^{2}} \frac{ V(\phi)}{V^{\prime}(\phi)} d\phi
	\end{equation}
	where $ \phi (t_{i} ) $ and $\phi (t_{f} ) $ are the values of the inflaton field at the time of Horizon crossing and at the end of inflation, respectively. At this point, we can start to discuss the implications of the swampland conjectures on this inflationary regime. A point to note is that for the dS conjecture is that if a low energy EFT scenario is consistent with either the original dS conjecture (2) or it's refined form (3), then it can have a consistent UV completion with regards to the dS criterion. Hence one can use only one of these conjectures in this regard if they like to, and we will be using the original dS conjecture (2) instead of the refined form in our analysis.  
	The issues of the swampland conjectures with single field inflation were firstly discussed in \cite{kinney2019zoo}, where the background cosmology was also general relativistic. Two of the very strong disagreements between the conjectures and inflation concerned the bounds on the $\epsilon$ parameter during inflation and the number of e-folds. In their work, they showed that the dS conjecture forbids the $\epsilon$ parameter to satisfy it's usual bound needed for sufficient inflation, which is $ \epsilon << 1 $. It is easy to see how this is the case, as this $\epsilon$ parameter bound for single field inflation in a GR based cosmology is given as \begin{equation}
	\epsilon = \frac{m_{p}^{2}}{2} \left( \frac{V^{\prime}}{V} \right)^2 \geq \frac{c^2}{2}
	\end{equation}
	where c is an $ \mathcal{O}(1) $ parameter in the original dS conjecture (2). The above equation makes it clear that it would be problematic to achieve $ \epsilon << 1 $ during inflation considering $ c \sim \mathcal{O} (1) $. From this definition, they were further able to work out some bounds on the scalar spectral index and they then found out that there is a disagreement between what the data on single field inflation and string theory itself predicts about the order estimates of the c parameter. Namely, the c parameter should be an $\mathcal{O}(0.1) $ term for it to be consistent with the data on inflation instead of $\mathcal{O}(1) $ as estimated from String theory.
	The next serious issue is concerns the e-fold number during Inflation. If one considers both the distance (1) and the dS conjectures (2) to hold true, then one finds out that these conjectures induce a fatal bound on the number of e-folds during inflation. This can also be readily shown here, as the number of e-folds for single field  (in a GR based cosmology) can be roughly written as \begin{equation}
	N \simeq \frac{\Delta \phi}{m_{p}	} \frac{1}{m_{p} \frac{V^{\prime}}{V} }
	\end{equation}
	Applying both the distance ($	\frac{\Delta \phi}{m_{p}} \leq d \sim \mathcal{O} (1) $ ) and dS conjecture( $	m_{p}  \frac{| V^{\prime} |}{V} \geq c \sim \mathcal{O} (1) $) here
	, one finds out that the e-fold number is constrained to be less than unity by these criterion. This is an incredibly distressing prediction, as the latest observational bounds \cite{akrami2020planck1} require at least around 60 e-folds in order for Inflation to explain the problems of big bang cosmology which it was originally brought in to rectify. Hence these explorations seem to suggest that all kinds of single field inflationary models in a GR based cosmology lie in the swampland and so, possibly, would not be viable with quantum gravity eventually. \\ \\
	However, now we make the case that single field inflation can be consistent with the swampland conjectures even when the background cosmology is described by GR, albeit with some Lorentz violations. We can now tackle both the prime concerns raised in \cite{kinney2019zoo} and find bounds on the parameter( $ \kappa $ due to Lorentz violation) which would allow us to rectify these issues. Considering the dS conjecture (2), one can write the following bound on the $ \epsilon $ parameter (19) as \begin{equation*}
	\epsilon \geq \frac{c^{2}}{2 (1 + \kappa)}
	\end{equation*}
	One can clearly notice that the dS conjecture bound for this $\epsilon$ parameter is similar to the one for the usual single field model encountered in (23), with the difference between both being the $ ( 1 + \kappa) $ term in the denominator here. As $ c \sim \mathcal{O} (1) $, in order for $ \epsilon << 1 $  one requires that \begin{equation}
	2 (1 + \kappa) >> 1 \
	\end{equation}
	This provides us a lower bound on the Lorentz violating parameter in order for the inflationary regime to be consistent with the swampland. Furthermore, we can roughly write the e-fold number(22) during inflation in this case as \begin{equation}
	N \simeq (1+ \kappa) \frac{\Delta \phi}{m_{p}	} \frac{1}{m_{p} \frac{V^{\prime}}{V} }
	\end{equation}
	Again applying the dS and distance conjectures together here, we find that the term $ \frac{\Delta \phi}{m_{p}	} \frac{1}{m_{p} \frac{V^{\prime}}{V} } $ has to be less than 1. Hence, in order to have sufficient inflation one requires \begin{equation}
	(1+ \kappa) >> 1   
	\end{equation}
	Hence considering the conditions in (25) and (27), the final bound on $\kappa$ is \begin{equation}
	    \boxed{\kappa >> 1}
	\end{equation} Hence in order for inflation to be consistent with the swampland conjectures in this regime, the minimum requirement that  needs to be satisfied is $ \kappa > 1 $. This will get us rid of the issues of both the $\epsilon$ parameter and the e-fold number and hence this lower limit on $ \kappa $ is what we need in order to have swampland consistent inflation in this regime. \\ \\ 
	A question one might wonder about now is, how would we actually get to know whether or not this inequality is satisfied in a particular inflationary model ? The answer to that lies in the perturbation parameters for inflation, in particular the scalar spectral index. The definition of the scalar spectral index remains the same here as it is for usual inflationary models \cite{almeida2017cosmology,baumann2009tasi} \begin{equation}
	n_{s} -1 = 2\eta - 6\epsilon
	\end{equation}
	One can even further work out other perturbation parameters like the tensor-to-scalar ratio, the tensor spectral index etc. but the scalar spectral index is all we need here to constrain $ \kappa $ in this case using the latest observational data from the Planck experiment \cite{akrami2020planck1}. In the next section we will now constrain the Lorentz violating parameter in 3 different inflationary models and try to ascertain whether the constraints from the swampland conjectures can be satisfied by a significant amount of potentials.
	
	\section{Analysing the swampland consistency of some models in the Lorentz violating regime }
	
	In this section, we will work out the Lorentz violating parameter in the Higgs, Radion Gauge and Spontaneous Symmetry breaking inflationary models and check whether these models are swampland consistent in a Lorentz violating regime. 
	
	\subsection{Higgs Inflation }
	
	Higgs Inflation is an inflationary regime of particular phenomenological interest \cite{bezrukov2008standard,bezrukov2009standard,bezrukov2009standard1,garcia2011higgs}. As the name suggests, here the well known Higgs field h itself is considered to be playing the role of the inflaton field. The well known inflaton potential in this case is  \cite{bezrukov2008standard,bezrukov2009standard,garcia2011higgs,martin2014encyclopaedia} \begin{equation}
	V(\phi) = M^{4} (1- e^{-\sqrt{2/3} \phi/m_{p} })^{2}
	\end{equation}
	where M is a mass scale (in fact this will remain the convention for mass scale depiction for the later potentials in this work as well). One interesting thing to note here is that the Higgs inflation potential is an example of an inflationary model with no free parameters. In order to obtain the $ \kappa $ values allowed by this model, we first need to find the slow roll parameters $ \epsilon $ and $\eta$ ( as defined in (19) and (20) ) for the potential (30),. Although this model has been shown to be consistent with observational data, it has been shown that even the higgs potential can be in tensions with the swampland conjectures in the usual GR inflationary regime \cite{denef2018sitter}. Hence it would be interesting to see whether considering a Lorentz violating regime for inflation would help alleviate these issues.\\ \\
	For this potential, one can find these parameters to be \begin{equation}
	\epsilon = \frac{4}{3 (1 + \kappa) \left(e^{\sqrt{2/3} \phi / m_{p}}-1\right)^2}
	\end{equation}
	\begin{equation}
	\eta = \frac{4 \left(2 - e^{\sqrt{2/3} \phi / m_{p}}\right)}{3 (1 + \kappa)
		\left(e^{\sqrt{2/3} \phi / m_{p}}-1\right)^2} 
	\end{equation}
	After being powered with the knowledge of these parameters, we can now find the scalar spectral index $ n_{s} $ after some algebra as \begin{equation}
	n_{s} -1 =  -\frac{4 (-1 + \coth[\phi/(\sqrt{6} m_{p})]) \coth[\phi/(\sqrt{6} m_{p})] }{ 3(1 + \kappa)}
	\end{equation}
	The latest observational constraints \cite{akrami2020planck1} on the scalar spectral index estimate that $ n_{s} = 0.9649 \pm 0.0042$ at the time of Horizon crossing, hence for our purposes here we can take $ n_{s} \simeq 0.9649 $ and be assured of considerable precision in our analysis. Using this value of the scalar spectral index, we can arrive at the following value for the Lorentz violating parameter \begin{equation}
	\kappa \simeq 40 \left(\coth \left(\frac{\phi}{\sqrt{6} m_{p}}\right)-1\right) \coth \left(\frac{\phi}{\sqrt{6}
		m_{p}}\right)-1
	\end{equation}
	We consider the $\phi$ value at the time of horizon crossing $ \phi = \phi (t_{i}) $  to be around the reduced planck mass  \begin{equation}
	\phi \simeq  m_{p}
	\end{equation} (We will be justifying this choice in the end of this subsection)Inserting this into the above equation, we finally find the Lorentz violating parameter to be \begin{equation}
	\kappa \simeq  154.495
	\end{equation}
	This is well above the minimum bound implied by the swampland conjectures on inflation in the Lorentz violating regime , which is $ \kappa > 1 $(28). To now show that one is not wrong to take $ \phi \simeq m_{p} $ at the time of horizon crossing, we take the help of the e-folding number(22) for which we first need to find $\phi$ at the end of inflation. This can be achieved quite easily using the fact that $ \epsilon = 1 $ at the end of inflation, which from (22) can be written as \begin{equation}
	\frac{4}{3 (1 + \kappa) \left(e^{\sqrt{2/3} \phi(t_{f}) / m_{p}}-1\right)^2} = 1 
	\end{equation} This is quite straightforward to solve and one  gets \begin{equation}
	\phi(t_{f}) \simeq 0.1084 m_{p}
	\end{equation} Now plugging the value of $\kappa$ (36) into  this $\phi$ at the end of inflation and using our estimate $\phi(t_{i}) \simeq m_{p}$, we can  compute the integral(22) for the Higgs Potential (30) to get the e-folding number as \begin{equation}
	N = \frac{1 + \kappa}{m_{p}^{2}} \int_{\phi(t_{f})}^{\phi(t_{i})} \frac{1}{2} \sqrt{\frac{3}{2}} m_{p} \left(e^{\frac{\sqrt{\frac{2}{3}} \phi}{m_{p}}}-1\right) \simeq 52
	\end{equation} This is well within the usual range between 50-70 for the e-fold number which is required for inflation to solve the problems of conventional big bang cosmology and  be consistent with the latest observational data \cite{akrami2020planck1,aghanim2018planck} and so the estimate $\phi \simeq m_{p}$ is justified and is not in any kind of tensions with the data.  This analysis tells us that Higgs Inflation is consistent with the swampland conjectures even in a GR based cosmology when one considers Lorentz violations in the cosmological background. It is also interesting to note that this is a zero parameter model (in the sense that the potential has no free parameters to tune accordingly) but is still consistent with the swampland conjectures. 
	
	\subsection{Radion Gauge Inflation }
	
	Radion Gauge inflation was first studied in \cite{fairbairn2003radion} and is an extension of the gauge inflation scenario in which the radius modulus field around which the Wilson loop is wrapped assists inflation as it shrinks \cite{freese1990natural}. The potential in this scenario can be written as \cite{martin2014encyclopaedia} \begin{equation}
	V(\phi) = \frac{M^4 (\phi^2/m_{p}^{2})}{\alpha + (\phi^2/m_{p}^{2})}
	\end{equation}  
	where M is again a mass scale and $\alpha$ is a positive dimensionless parameter, which would act as the only free parameter of the model. This potential can also be found in S-dual superstring models \cite{de1996inflation} Although Radion gauge inflation has never been studied in the context the swampland for GR based single field inflation, the issues pointed out for GR based inflation in \cite{kinney2019zoo} are so strong that they would generally rule all kinds of GR based potentials in the swampland and hence, it would be interesting to see how this model stands with these criterion in a Lorentz violating regime. \\ \\
	We proceed similar to the way we did for Higgs Inflation in the subsection 4.1, by firstly finding out the slow roll parameters in this regime. One can find that the parameters, after some algebra, take the forms \begin{equation}
	\epsilon = \frac{2 \alpha^2 m_{p}^6}{(1 + \kappa) \left(\alpha m_{p}^2 \phi+\phi^3\right)^2}
	\end{equation}
	\begin{equation}
	\eta = \frac{2 \alpha m_{p}^4 \left(\alpha m_{p}^2-3 \phi^2\right)}{(1 + \kappa) \left(\alpha m_{p}^2 \phi+\phi^3\right)^2}
	\end{equation}
	After finding the slow roll parameters, it is straightforward to write the scalar spectral index using its definition (29) and one finds the index to be \begin{equation}
	n_{s} - 1 = - \frac{4 \alpha m_{p}^4 \left(2 \alpha m_{p}^2+3 \phi^2\right)}{(1 + \kappa) \left(\alpha m_{p}^2 \phi+\phi^3\right)^2}
	\end{equation} 
	Again using the estimate $ n_{s} \simeq 0.9649 $ and  the field value at Horizon crossing to be $ \phi \simeq m_{p} $ (we will again justify this towards the end of this as in subsection 4.1), one finds the following relation for the Lorentz violating parameter in terms of $\alpha$ as \begin{equation}
	\kappa = \frac{\alpha (227.92 \alpha+341.88)}{(\alpha+1)^2} - 1 
	\end{equation}
	In order to be consistent with the swampland conjectures, one requires the minimum bound $ \kappa >1 $ on the Lorentz violating parameter and this implies \begin{equation}
	\kappa = \frac{\alpha (227.92 \alpha+341.88)}{(\alpha+1)^2}-1 > 1 \implies \alpha > 0.0058
	\end{equation}
	The requirement imposed on the free parameter $\alpha$ in order for this model to be swampland consistent is quite minimal and extremely in line with the basic view that $\alpha$ is a positive parameter. Before concluding, we again justify the choice we made for $ \phi(t_{i}) \simeq m_{p} $ by computing the e-fold number. As in subsection 4.1, we firstly find the $\phi$ value at the end of inflation using $\epsilon = 1 $, which gives us \begin{equation}
	\frac{2 \alpha^2 m_{p}^6}{(1+\kappa) \left(\alpha m_{p}^2 \phi+\phi^3\right)^2} = 1 
	\end{equation} Solving this analytically for $\phi$ can be quite cumbersome but fortunately, we can do this in an easier way. Given the definition of $\kappa$ (44) in terms of $\alpha$, we can plug some particular values of $\alpha$ , thereby obtaining the corresponding $\kappa$, and then use those values to obtain the $\phi$ value at the end of inflation and compute the Number of e-folds. For simplicity, let's take $\alpha = 1 $, which would result in $\kappa = 141.45$. Using these values in (46), one can find the $\phi$ value at the end of inflation to be \begin{equation}
	\frac{0.01404 m_{p}^6}{\phi(t_{f})^2 \left(m_{p}^2+\phi(t_{f})^2\right)^2} = 1 \implies \phi(t_{f}) \simeq 0.1143 m_{p}
	\end{equation} Using the $\phi(t_{f})$ to be the above value and $\phi(t_{i}) \simeq m_{p} $ one finds the e-folding number to be \begin{equation}
	N = \int_{\phi(t_{f})}^{\phi(t_{i})}  \frac{71.225 \phi \left(m_{p}^2+\phi^2\right)}{m_{p}^4} d\phi \simeq 53
	\end{equation}  We see that even for an elementary value of $\alpha = 1$, one can  sufficient number of e-folds of Inflation needed for solving the conventional problems of big bang cosmology and being consistent with the latest observational data \footnote{ We understand that a reader might be tempted to tend towards a more graphical approach here, which is that one might want to make a graph of N vs $\alpha$ and would want to look for the required e-fold consistent $\alpha$ by doing that rather than plugging values on their own. Even we would have preferred to go that route but as it turns out, analytically finding N as a function of $\alpha$ for arbitrary values of the parameter is not quite feasible, partly because changing the $\alpha$ values also changes the $\phi(t_{f})$ value significantly(which again is quite hard to solve for arbitrary values) and this also, hence, changes the integration limits for computing the e-folding number. This was the reason we decided to plug in $\alpha$ values ourselves to do this analysis, and we were still able to show that this model can nevertheless be easily consistent with the observational requirements for the e-folding number. } \cite{akrami2020planck1,aghanim2018planck}. Hence considering $\phi(t_{i}) \simeq m_{p} $ would not be a wrong choice in this scenario and is completely consistent with both the swampland conjectures and the data. This analysis finally allows us to conclude that Radion gauge inflation can be easily viable with the swampland conjectures in a Lorentz violating regime even when the background cosmology is general relativistic.

	\subsection{Spontaneous Symmetry Breaking Inflation }
	Spontaneous symmetry breaking inflation, in the sense that we are going to be discussing here, was firstly worked on by Moss in \cite{moss1985primordial}. Moss considered the forthcoming potential in the framework of models with spontaneous symmetry breaking where $\phi $ represents one of the components of a Higgs field. The potential in this inflationary regime takes the form \begin{equation}
	V(\phi) = M^4 \left(1 + \frac{\alpha \phi^2}{m_{p}^2}+\frac{\gamma \phi^4}{m_{p}^4} \right)
	\end{equation}
	where $\alpha$ and $\gamma$ are constant dimensionless parameters and M is again a mass scale. This model is different from the others we have discussed so far in that it has 2 free parameters in $\alpha$ and $\gamma$ instead of just one (like the radion gauge model) or none (like Higgs Inflation). It is interesting to mention that besides spontaneous symmetry breaking, this type of potential also finds place in certain SUSY and Spontaneous D-Parity Breaking inflationary scenarios\cite{dine1997inflaton,riotto1998inflation,gong2008inflation}. It also becomes interesting to see the consistency of this model with the swampland as spontaneous symmetry breaking is the way in which Lorentz violating effects are introduced in the Standard Model extension \cite{kostelecky1989spontaneous}, which is an EFT framework which contains the Standard Model, General Relativity and other possible operators which can break Lorentz symmetry.\\ \\
	The slow roll parameters for this potential take the form \begin{equation}
	\epsilon = \frac{2 \left(\alpha m_{p}^3 \phi+2 \gamma m_{p} \phi^3\right)^2}{(1 + \kappa) \left(\alpha m_{p}^2 \phi^2+\gamma \phi^4+m_{p}^4\right)^2}
	\end{equation} 
	\begin{equation}
	\eta = \frac{2 m_{p}^2 \left(\alpha m_{p}^2+6 \gamma \phi^2\right)}{(1 + \kappa) \left(\alpha m_{p}^2 \phi^2+\gamma \phi^4+m_{p}^4\right)}
	\end{equation}
	This allows us to find the scalar spectral index (29) to be \begin{equation}
	n_{s} -1 = \frac{4 m_{p}^2 \left(-2 m_{p}^4 \phi^2 \left(\alpha^2-3 \gamma\right) \right)}{(1 + \kappa) \left(\alpha m_{p}^2 \phi^2+\gamma \phi^4+m_{p}^4\right)^2} - \frac{5 \alpha \gamma m_{p}^2 \phi^4 - \alpha m_{p}^6 + 6 \gamma^2 \phi^6\ }{(1 + \kappa) \left(\alpha m_{p}^2 \phi^2+\gamma \phi^4+m_{p}^4\right)^2}    
	\end{equation}
	\\
	We again take note of the fact that the index is being evaluated at horizon crossing, and  consider $ \phi = m_{p} $ which for this particular potential can be considered pretty natural as spontaneous symmetry breaking inflationary models lie in the class of small field models and the field value doesn't take values much higher than the Planck mass but we will still be justifying this estimate later on. Using the estimate $ n_{s} \simeq 0.9649 $ again and imposing the minimum requirement for swampland consistency for Lorentz violating inflation, $\kappa > 1 $,  we get the following inequality \begin{equation}
	\kappa = \frac{683.76 \gamma^2+\gamma (569.8 \alpha-683.76)+\alpha (227.9 \alpha-113.96)}{(\gamma+\alpha+1)^2}-1 > 1 
	\end{equation}
	One can check with the help of any mathematical computation tool that this inequality can be satisfied for a huge range of $\alpha $ and $\gamma $, where both of them can be either positive or negative. Just for an example, we highlight one possible solution of the above inequality which puts the following simple bounds on $\alpha$ and $\gamma$,    \begin{equation}
	\alpha \leq -1 , \gamma < 0 
	\end{equation} 
	These restrictions on $\alpha$ and $\gamma $ do not represent a great deal of fine tuning and are just one of the many possible range of values which can be attained by these parameters in order for satisfy the swampland constraint for Lorentz violating inflation. In order to compute the e-folding number in this case, we can again go the route \footnote{We again choose this method instead of doing the forthcoming analysis in a fully analytical sense for arbitrary values of the free parameters for similar reasons as to why we chose a similar route for doing the e-folding analysis for radion gauge inflation in subsection 4.2 }  that we explored in the previous subsection and start by getting a particular value for $\kappa$ .In this case that can be obtained by choosing any $\alpha$ and $\gamma$ that satisfy (54), where just for simplicity we choose $\alpha = \gamma = -10$ . To get $\phi(t_{f})$ we again set $\epsilon = 1 $, which gives us \begin{equation}
	\frac{0.441 \left(m_{p}^3 \phi(t_{f})+2 m_{p} \phi(t_{f})^3\right)^2}{\left(m_{p}^4-10 m_{p}^2 \phi(t_{f})^2-10 \phi(t_{f})^4\right)^2} = 1 
	\end{equation}  
	The equation above can be solved to obtain $\phi(t_{f}) \simeq 0.271041m_{p}$. Using this value and our estimate $\phi(t_{i} ) \simeq m_{p}$(alongside our chosen values for $\alpha,\gamma $ and $\kappa $), we get the number of e-folds for the potential (49) by the integral (22) as \begin{equation}
	N = \int_{\phi(t_{f})}^{\phi(t_{i})} \frac{+216.24 m_{p}^2 \phi^2+216.24 \phi^4-21.624 m_{p}^4}{m_{p}^4 \phi+2 m_{p}^2 \phi^3} \simeq 60
	\end{equation}
	The number of e-folds of inflation produced in this case are again enough to solve the conventional problems of big bang cosmology and be consistent with the latest observational data \cite{akrami2020planck1,aghanim2018planck}. Hence our horizon crossing field estimate $\phi(t_{i}) \simeq m_{p}$ produces adequate inflation for such elementary values for the parameters $\alpha$ and $ \gamma$ whilst still being consistent with the data. This analysis finally allows us to conclude single field inflation can still be consistent with the swampland conjectures in an essentially GR based cosmology in a Lorentz violating regime.

	\section{Concluding remarks and discussion}
	To summarize, in this work we attempted to find a way in which single field inflationary models can be consistent with the swampland conjectures even when the cosmological background in essentially general relativistic. Although there has been significant work on this in recent times, the novelty of our current work lies in the fact that we achieve this goal in quite a trouble-free way by considering a time-like Lorentz violating background. We start off our work by firstly reflecting on the problems between the swampland conjectures and single field inflation,  which are particularly in unavoidable loggerheads in a general relativistic cosmology. We then discuss some crucial aspects of the Lorentz violating cosmology we have considered in our work, showing how the inflationary dynamics is affected by considering the Lorentz violations. We then show what is the requirement for Lorentz violating single field models to be consistent with the swampland conjectures, which eventually turns out to be a minimum bound on the value of the Lorentz violating. We then briefly touch upon the fact that quadratic inflation is not consistent with the swampland bound on the Lorentz violating parameter, building on previous work on the same model in this regime. After this we consider 3 inflationary models of deep phenomenological interest in the form of Higgs inflation, Radion gauge inflation and Spontaneous symmetry breaking inflation and show that all three of these models(which would otherwise have faced the same difficulties with the swampland in a simple GR based cosmology as their other compatriots)  are quite easily consistent with the swampland criterion in our Lorentz violating GR based cosmology. \\ \\
	The main takeaway from this work is that there can still be ways for (cold )single field inflation to be consistent with the swampland conjectures even when the background cosmology is essentially GR based. Another interesting outcome idea that can be pondered upon from our work here concerns the significance of Lorentz violations in the Early Universe. Lorentz symmetry is a cornerstone idea of Relativity and if essentially GR based inflationary regimes, which is in quite unavoidable tension with the swampland criterion, becomes rather easily consistent with these conjectures only by considering a certain form of Lorentz violation in the background cosmology then it could possibly have interesting implications from a quantum gravity point of view. The premise of the swampland conjectures is that these criterion are supposedly necessary conditions in order for low energy EFT's to have consistent UV completion, so the fact that a Lorentz violating cosmology makes it more tranquil for inflationary regimes to be consistent with the swampland could really give heed to the notion that quantum gravity points towards  Lorentz violations being more significant in the early universe than what one presumes till now . Also, it would be interesting to see in future works about the implications of a Lorentz violating scenario on warm inflation, non-GR based inflation and multi-field inflation in the context of the swampland and any corresponding implications it could have on more late universe scenarios like dark energy, dark matter and the Hubble tension.
	 
	\section{acknowledgments}
	
	The author would like to thank the Prof. Ralf Lehnert and all the organizers of the $ 4^{th} $ IUCSS Summer School on Lorentz- and CPT- Violating Standard Model extensions, which was hosted by Indiana University, Bloomington. The ideas for this work developed during the summer school and the author would like to express his deepest gratitude to the school organizers for putting together such an intellectually enriching event.
	
	\section{Data availability statement}
	
	There are no new data associated with this work.

	\bibliographystyle{unsrt}
	\bibliography{JSPJMJ1.bib}

\end{document}